%% file: ms.tex
\documentclass[a4paper,fleqn,usenatbib]{mnras}

\usepackage{newtxtext,newtxmath}

\usepackage[T1]{fontenc}
\usepackage{ae,aecompl}

\usepackage{graphicx}	
\usepackage{amsmath}	
\usepackage{amssymb}	

\usepackage{txfonts}
\usepackage[]{natbib}
\usepackage{lscape}
\usepackage{longtable}

\newcommand{\ebv}{$ E \left( B-V \right)$}
\newcommand{\ebvfg}{$ E \left( B-V \right)_{fg}$}

\newcommand{\evi}{$ E \left( V-I \right)$}
\newcommand{\MV}{$M_V$}
\newcommand{\vrad}{$\rm v_{rad}$}
\newcommand{\kms}{$\rm km \, s^{-1}$}

\newcommand{\Msun}{$\rm \,  M_{\odot}$}
\newcommand{\Osun}{$\rm \,  O_{\odot}$}
\newcommand{\Fesun}{$\rm \,  Fe_{\odot}$}
\newcommand{\Zsun}{$\rm \,  Z_{\odot}$}

\newcommand{\Mini}{\mbox{$M_{ini}$}}

\newcommand{\hb}{$ H {\beta}$}
\newcommand{\hg}{$ H {\gamma}$}

\title[First massive stars in SagDIG]{Massive stars in the
       Sagittarius Dwarf Irregular Galaxy \thanks{
  Based on observations made with the Gran Telescopio Canarias (GTC), 
  installed in the Spanish Observatorio del Roque de los Muchachos 
  of the Instituto de Astrof\'{\i}sica de Canarias, on the island of La Palma. 
       }
  }

\author[M. Garcia]{Miriam Garcia$^{1}$\thanks{E-mail: mgg@cab.inta-csic.es}
\\
$^{1}$Centro de Astrobiolog\'{\i}a (CSIC-INTA). Crtra. de Torrej\'on a Ajalvir km 4.
      28850 Torrej\'on de Ardoz (Madrid), Spain \\ 
}

\date{Accepted XXX. Received YYY; in original form ZZZ}

\pubyear{2017}

\begin{document}
\label{firstpage}
\pagerange{\pageref{firstpage}--\pageref{lastpage}}
\maketitle

\begin{abstract}
  Low metallicity massive stars hold the key to interpret numerous processes in
  the past Universe including re--ionization, starburst galaxies,
  high-redshift supernovae and $\gamma$--ray bursts.
  The Sagittarius Dwarf Irregular Galaxy (SagDIG, 12+log(O/H)=7.37) represents an important landmark
  in the quest for analogues accessible with 10-m class telescopes.
  This paper presents low-resolution spectroscopy
  executed with the Gran Telescopio Canarias
  that confirms that SagDIG hosts massive stars.
  The observations unveiled three OBA--type stars and one red supergiant candidate.
  Pending confirmation from high-resolution follow-up studies,
  these could be the most metal-poor massive stars of the Local Group.
\end{abstract}




\section{Introduction}
\label{s:intro}

Understanding the evolution of very metal-poor massive stars
is key to interpret star formation and feedback at previous Cosmic epochs,
and a necessary step towards the physics of the first stars of the Universe.
This has stimulated an enthusiastic effort to discover
resolved massive stars in environments
of decreasing metal content,
mostly dwarf irregular galaxies of the Local Group.

The Small Magellanic Cloud (SMC) is the current standard
for the low-Z regime,
but metal-poorer massive
stars are now at reach with multi-object spectrographs at 8--10m telescopes.
First quantitative analyses were directed to the 
IC~1613, NGC~3109 and WLM galaxies \citep{Bal07,Hal10,Tal11}
with encouragingly
low 1/7\Osun~ oxygen abundances, 
but subsequent studies indicate that their iron content is SMC-like ($\sim$1/5\Fesun) instead
\citep[][]{Gal14,Hal14,Bal15}.
Following efforts were directed to Sextans~A  \citep[][]{Cal16}
with both 1/10\Fesun~ and 1/10\Osun~ \citep[][]{KVal04}.
However, this value is still far from the $\sim$1/30\Zsun~   
average metallicity of the Universe at the peak of star formation history 
and from measurements of the inter-galactic medium at higher redshifts \citep{MD14}.

The Sagittarius Dwarf Irregular Galaxy (\textit{aka} Sgr dIG, 
ESO 594-4, UKS 1927-17.7; SagDIG from now on)
makes a promising environment
to find very metal-poor, resolved massive stars.
Its intermediate-age population
stands out as one of the most Fe-poor among the Local Group galaxies
that sustain star formation
\citep[e.g.][]{McC12},  
with values inferred from the location of
the red giant branch (RGB) that range from [Fe/H]=$-$2.0 to $-$2.45 \citep{Becal14,KAM99}.
Regarding the youngest populations,
the oxygen abundance determined from
the only confirmed \ion{H}{ii} region SagDIG-HII\#3  \citep{SHK91}
is also remarkably low
12+log(O/H) =7.37$\rm ^{ + 0.13}_{ - 0.11}$~ 
\citep{SIal02}.
Scaled by the Solar photospheric abundance \citep{Aal09},
this value is equivalent to 1/21\Osun.

While SagDIG is considered one of the extremely metal-poor galaxies of the Local Universe,
rivaling I~Zw18 \citep{FSM15},
it has never been explored for massive stars.
This paper 
presents the first spectroscopic observations of young massive stars in the galaxy.
Section~\ref{s:redu} describes the observations and data reduction.
Target identification, spectral types and membership
are assessed in Section~\ref{s:SpT}.
The discovered red supergiant candidate is discussed in the context
of the Local Group in Section~\ref{s:RSGZ}.
Finally, summary and some prospects for future work are provided in Section~\ref{s:fin}.

\begin{figure}
\centering
   \includegraphics[width=0.5\textwidth]{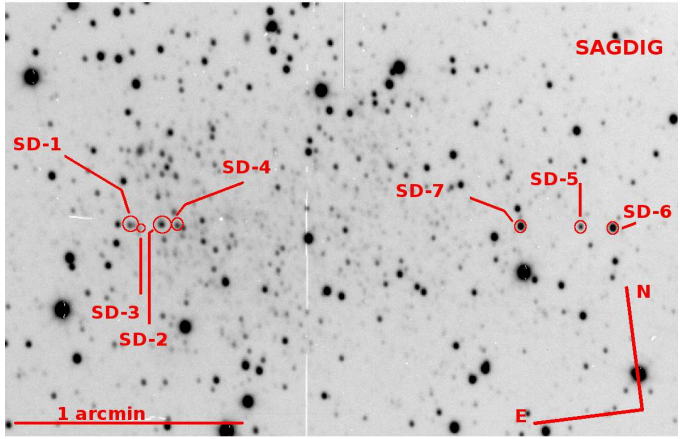}
   \caption{
     SagDIG: GTC--OSIRIS acquisition image (1.2 arcsec seeing, filter--free), and
     programme stars.
}
   \label{F:chart}
\end{figure}

\begin{figure*}
\centering
   \includegraphics[width=\textwidth]{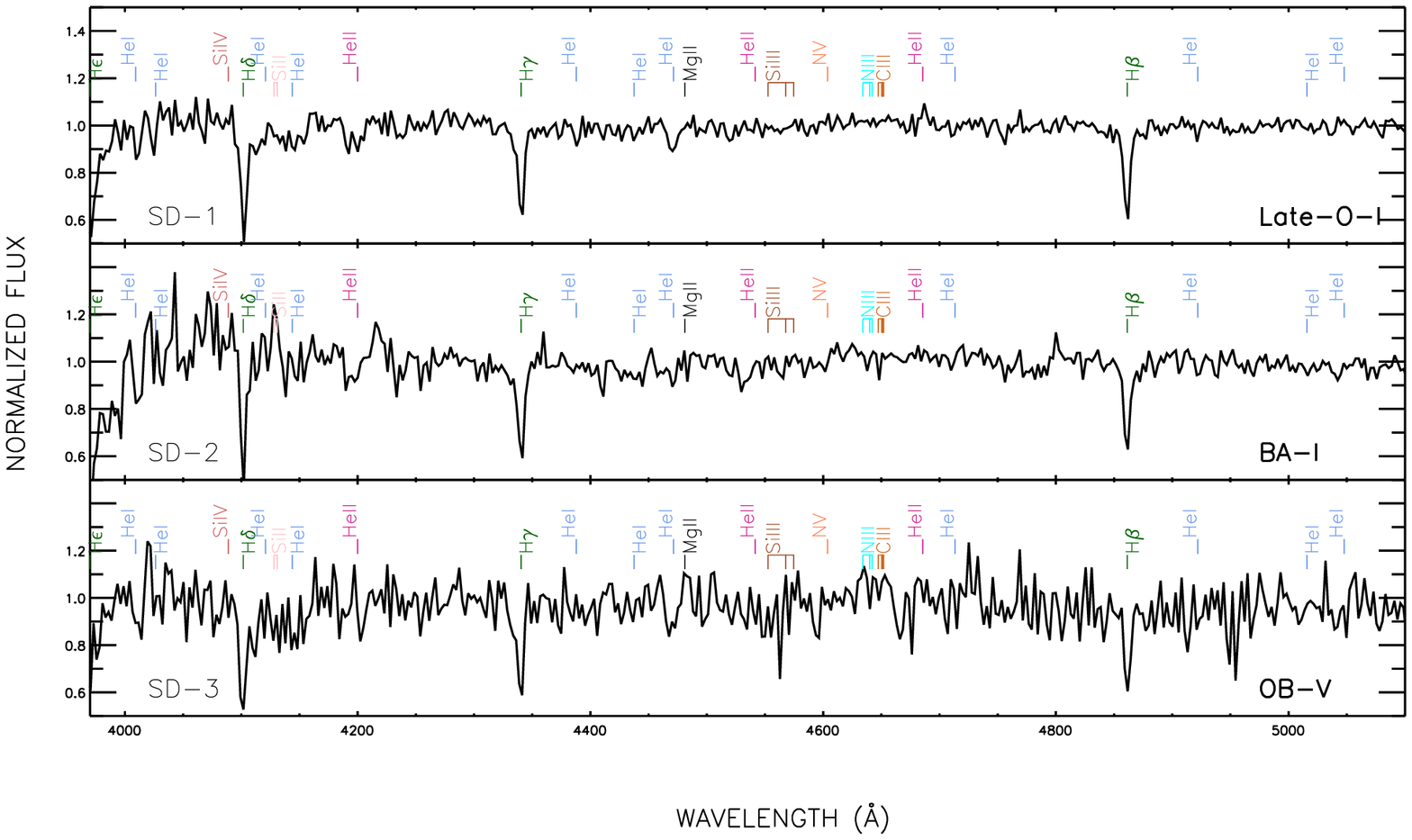}
   \caption{GTC--OSIRIS spectra of blue massive stars in SagDIG. The data have been rebinned
     to 3 pixels for clarity, and corrected by heliocentric and radial velocity.
     The three spectra show a clear Balmer series.
}
   \label{F:sp}
\end{figure*}

\section{Data and reduction}
\label{s:redu}

The observations were executed with the OSIRIS spectrograph 
at the 10.4-m Gran Telescopio Canarias (GTC), programme GTC67-13A (PI M. Garcia).
The 1.2 arcsec slit and R2000B VPH were used
to achieve resolution $R \sim 1000$~ at $\sim$4000--5500\AA.
The slit was aligned on targets SD-1 and SD-7
to cross the galaxy (Fig.~\ref{F:chart}) and enclose several UV--bright sources.
A total of 2 hours of grey sky, 1.2 arcsec seeing exposure time were devoted to the slit.

The data were reduced following standard \textsc{iraf}\footnote{
\textsc{iraf} is distributed by the National Optical Astronomy Observatory, operated by the 
Association of Universities for Research in Astronomy (AURA) under agreement with the National Science Foundation.
}
procedures adapted for the team's OSIRIS runs \citep{GH13a}.
The spectra were flux--calibrated 
against the flux standard GD190,
achieving about 5 per cent accuracy
for $\lambda > $4250\AA.
Finally, the spectra were set to the heliocentric frame and
corrected by radial velocity.
The latter were calculated
from the Doppler shift measured at the core of the Balmer lines.
The reduced spectra are shown in Figs.~\ref{F:sp}-\ref{F:sg}.

The programme faced the challenges related to observations of faint, distant stars:
contamination by nearby sources and small number of counts over the background level.
For instance the spectrum of SD-1,
with V$\sim$20.2, registered a factor $\sim$2.5 less counts than
the extracted sky spectrum.
The resulting signal to noise ratio (SNR) is consequently poor
even after careful sky subtraction
and some Solar features remained.
These shortcomings discourage future optical range spectroscopic observations
of targets fainter than V=20 mag in grey time.

Targets SD-1 to 4
are found in an area of extended continuum emission
\citep[detected in Far--Ultraviolet imaging as well, e.g.][]{Moal14}
that may contaminate the extracted spectra.
In order to minimize blends,
the sources were extracted 
discarding the wings of the spatial distribution of photons in the 2-D images.

\input{star_phot.tex}    

\section{Source identification, spectral classification and membership}
\label{s:SpT}

SagDIG is projected close to the Galactic Centre
and the line of sight is severely contaminated with Milky Way stars. 
Hence, ensuring that the programme targets belong to the galaxy was 
an issue. 
This section discusses jointly target identification, spectral classification,
membership, and their consistency.
The location of the programme stars in the colour--magnitude diagram (CMD) is
evaluated as a final check.

The programme targets were identified in the
\citet[][hereafter B14]{Becal14} and \citet[][M14]{Moal14} catalogues
by comparing the GTC acquisition images against
HST--ACS archival observations (GO-9820 P.I. Y. Momany).
Identification numbers and photometry are listed in
Table~\ref{T:phot}.

\begin{figure}
\centering
   \includegraphics[width=0.5\textwidth]{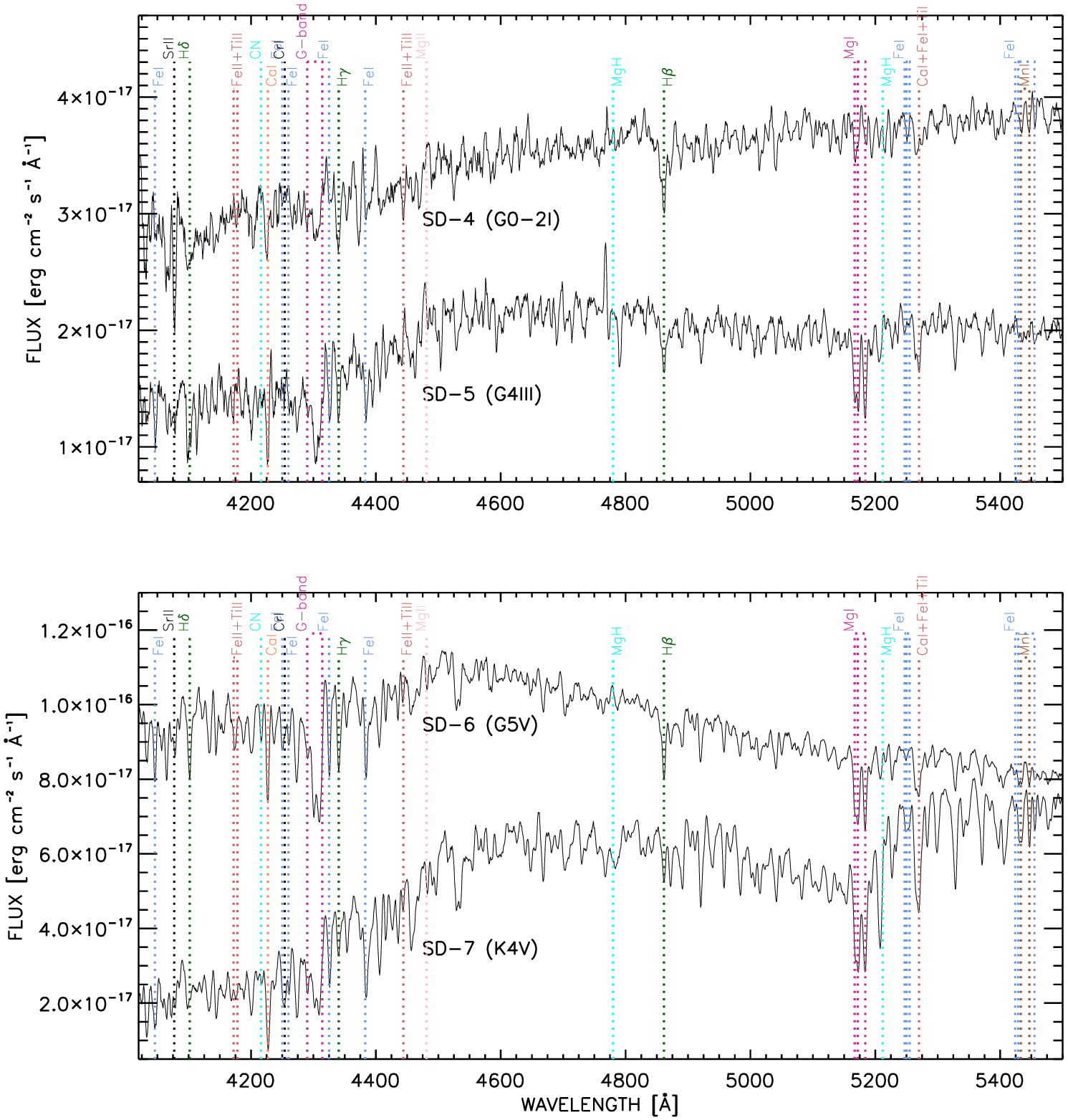}
   \caption{Red programme stars.
     The flux--calibrated spectra were smoothed to 5 points,
     and corrected by heliocentric and radial velocity.
     In both pannels the upper spectrum
     has been shifted by a constant for clarity
     (SD-4: 1.5E$-$17 $\rm erg \, cm^{-2} \, s^{-1} \, \AA^{-1}$, SD-6: 3E$-$17 $\rm erg \, cm^{-2} \, s^{-1} \, \AA^{-1}$).
}
   \label{F:sg}
\end{figure}

The spectral classification of blue stars,
whose most prominent spectral feature
is the Balmer series, was carried out following classical
criteria \citep[see e.g.][]{GH13a}.
Their spectra are shown in Fig.~\ref{F:sp}.
The rest of the sample (Fig.~\ref{F:sg}) was classified after  \citet{D16}'s scheme
for cool stars (G--types and later)
in the LMC and the SMC,
and the spectral classification atlas by Gray (2000)\footnote{
  \texttt{https://ned.ipac.caltech.edu/level5/Gray/frames.html}}.
They were also compared against SMC and MW standards.

Membership of the programme stars with counterpart in
M14's catalogue, that registered
relative shifts from  multi-epoch HST visits,
was ensured by the lack of proper motions.
In addition, the estimated absolute magnitudes were 
compared against the calibrated values for the spectral types assigned in this work.
The consistency of the stellar radial velocities and SagDIG's systemic velocity
\citep[$-$78.5 \kms,][]{McC12}
was not determinant because of the low spectral resolution of these observations
($\sim$ 300 \kms, hence typical \vrad~ uncertainties of $\sim$ 100 \kms).

The following subsections provide
more details on the targets that were positively identified
as SagDIG massive stars.

\subsection{SD-1}

The HST--ACS images reveal that SD-1
is the blend of two stars $\rm \sim$ 1.1 arcsec apart,
SD-1-blue (M14-33466; B14-8903, V=20.190)
and SD-1-red (B14-8909, V=21.774).
SD-1 will be identified with the brightest star of the blend, SD-1-blue, from now on.

The Balmer series is clearly detected in its spectrum (Fig.~\ref{F:sp}), and the comparison against the 
sky spectrum confirms that the stellar component dominates \hg~ and \hb.
The data also display \ion{He}{i}+\ion{He}{ii}~4026\AA, \ion{He}{i}~4387\AA~ and \ion{He}{i}~4471\AA,
but the detection of \ion{He}{ii} lines is unclear.
\ion{He}{ii}~4541\AA~ is absent and the
\ion{He}{ii}~4200\AA~ and \ion{He}{ii}~5411\AA~ absorptions 
overlap with sky features.

The exception is \ion{He}{ii}~4686\AA, detected in emission.
Its width together with the absence of \ion{H}{i} and \ion{He}{i} emission lines
renders a nebular origin unlikely,
and the comparison against the subtracted sky spectrum discards the feature as an artefact.
The  \ion{He}{ii}~4686\AA~ emission could be consistent with a late--O type,
with the rest of \ion{He}{ii} lines too weak for detection at the SNR of the spectrum.
The non--detection of \ion{Mg}{ii}~4481\AA~ and \ion{Si}{iii}~4552\AA~ supports the late--O classification,
although it could also be caused by the low metallicity of the star.

The spectrum of SD-1 is reminiscent of star C1\_31 in NGC~55 \citep{Cal08}.
C1\_31 is mostly featureless except for the Balmer lines and 
a prominent broad \ion{He}{ii}~4686\AA~ emission, with \ion{He}{i}~4471\AA~ clearly seen but no \ion{He}{ii}~4541\AA.
It was classified as early--O supergiant.
Similarly, SD-1 is classified as late--O~I, with the \ion{He}{ii}~4686\AA~ emission
and narrow Balmer lines supporting the supergiant luminosity class.
The estimated absolute magnitude \MV=$-$6.1 agrees with
the calibrated value for late--O supergiants \citep{M98}.

\subsection{SD-2}

SD-2 is the blend of M14-31722 (F606W=20.278) and M14-31570 (F606W=21.839),
0.7 arcsec apart and unresolved.
SD-2 is identified with the brightest source M14-31722.

Its spectrum shows the Balmer series,
\ion{He}{i}~4471\AA, \ion{He}{i}~4387\AA~ and \ion{He}{i}~4922\AA~ (Fig.~\ref{F:sp}).
There may be some \ion{Mg}{ii}~4481\AA,
but no lines of \ion{Si}{iii}, \ion{Si}{iv} or \ion{He}{ii}.
The calibrated stellar flux corresponds to a colder object
than SD-1.  
Since the width of the Balmer lines is similar to SD-1,
and the strength of the \ion{He}{i} lines is almost at noise level,
SD-2 is classified as a BA-supergiant.
The estimated absolute magnitude \MV=$-$5.1
is consistent with a blue
supergiant located at SagDIG.

\subsection{SD-3}

Target SD-3 is faint and 
off--slit, resulting in extremely low spectral SNR.
The sky contribution 
amounts to 83 per cent of the total flux at blue wavelengths, increasing to 91 per cent at 4800\AA.

SD-3 overlaps with the red component of the SD-1 blend.
However, there is a shift between SD-3's strongest features
and the same lines in SD-1
reflecting the different radial velocities of the two stars
and confirming that the lines genuinely belong to SD-3.

The spectrum of SD-3 (Fig.~\ref{F:sp}) exhibits the Balmer series and \ion{He}{i}~4471\AA.
The \ion{He}{ii} lines, but also \ion{Si}{iii}~4552\AA~ and \ion{Mg}{ii}~4481\AA, are absent
suggesting late--O or early--B spectral type. 
The spectrum lacks any \ion{He}{ii}~4686\AA,
but contamination by SD-1-red begins to be significant in this region.
The wings of \hg~ are dominated by the stellar component,
and the profile is wider than SD-1's, suggesting class III or V.
Since the absolute magnitude agrees with \citet{M98}'s calibration for O9--B0~V (\MV=
$-$4.4 to $-$3.8),
SD-3 is classified as an OB-dwarf.

\subsection{SD-4}

SD-4 (M14-30481, F606W=20.503)
is contaminated by a nearby red star with similar magnitude (M14-30025, F606W=20.360).
According to their proper motions, both belong to SagDIG.

Luminosity class was constrained in the first place, following \citet{D16}.
Because the 5167\AA~ component is the strongest of
the magnesium triplet \ion{Mg}{i}~5167,~5172,~5184\AA,
SD-4 was classified a supergiant.
The relative strength of \ion{Fe}{i}~5247,~5250,~5255\AA~
\textit{vs} \ion{Ca}{i}+\ion{Fe}{i}+\ion{Ti}{i}~5270\AA, and \ion{Mn}{i}~5433\AA~ vs \ion{Mn}{i}~5447\AA~
support this luminosity class.

The presence of the G--band and its relative strength compared to \hg~
and \ion{Ca}{i}~4227\AA~
suggests that the spectral type is slightly later than G0.
The relative strength of \ion{Fe}{i}~4325\AA~ compared to both \hg~ and \ion{Fe}{i}~4383\AA, on the other hand,
suggests spectral type G0 or earlier.
A G0--2~I type is finally assigned.

It is noted that the estimated absolute magnitude \MV=$-$5.3
is slighlty fainter than the calibrated \MV=$-$6.4 G0~I.
Nevertheless, SD-4's membership to SagDIG is supported
by the lack of detection of proper motions by M14.
Since red supergiants (RSG) stars may experience spectral
variability
(see Sect.~\ref{s:RSGZ}),
SD-4 is proposed as a RSG-candidate in SagDIG.

\begin{figure}
\centering
   \includegraphics[width=0.5\textwidth]{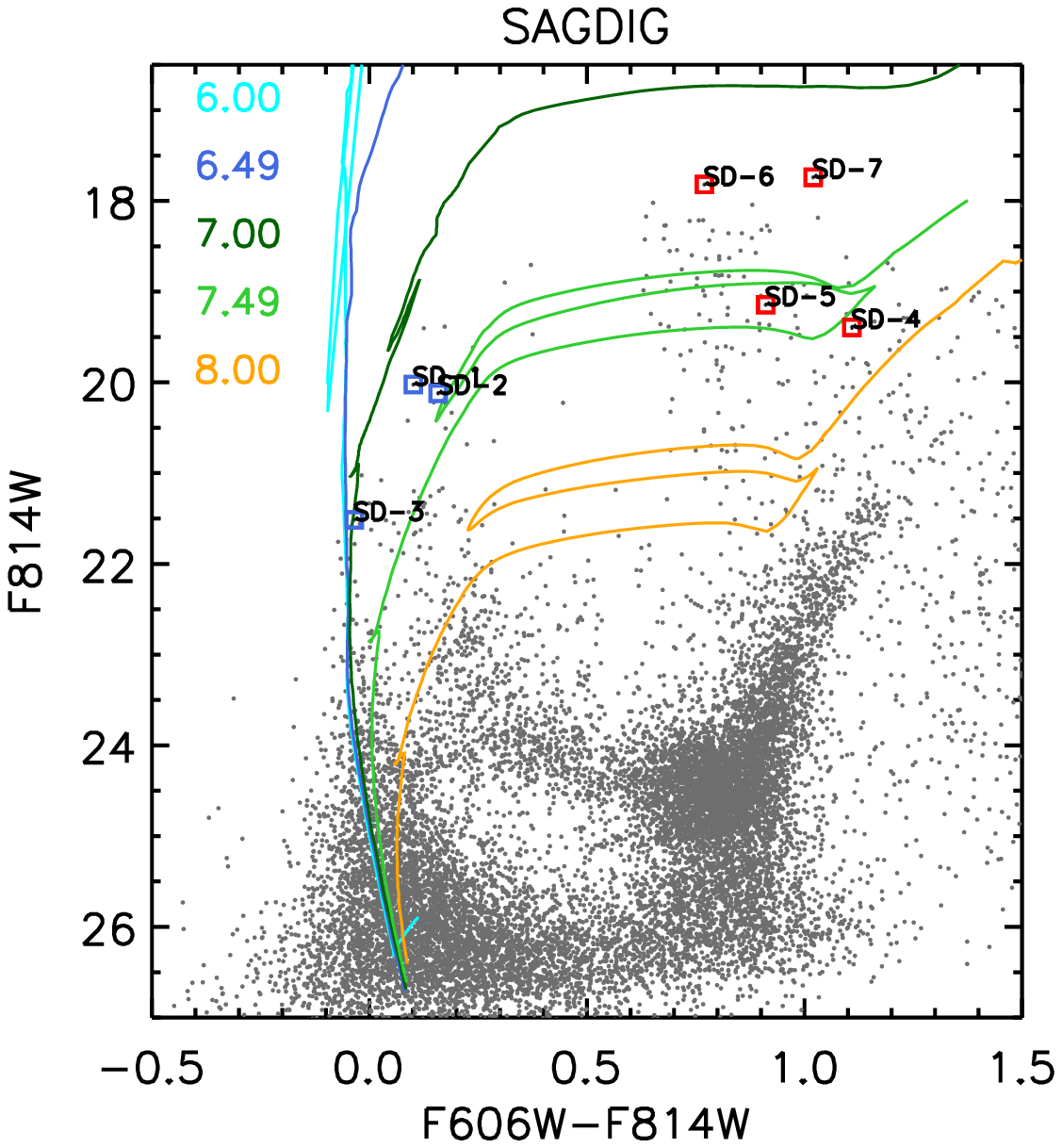}
   \caption{
     SagDIG's colour--magnitude diagram built from M14's catalogue (grey dots).
     \citet{MGB17}'s isochrones for Z=0.001 ($\sim$1/15\Zsun) are included for comparison,
     with colour coding $ \log \left( age[yrs] \right) $=6.0,6.49,7.0,7.49,8.0.
     Younger isochrones are not included because
     they overlap at F606W$-$F814W$\sim$0 in the F814W=20--22 range.
     The isochrones were shifted by SagDIG distance modulus $\left( m-M \right) _0 $=25.10
     and reddened by the equivalent to \ebv=0.22
     ($A_{F606W} \,$= 0.646, $A_{F814W} \,$= 0.410, after \citet{SJB05}).
}
   \label{F:EBVHII}
\end{figure}

\subsection{Target stars in the colour--magnitude diagram}
\label{s:cmd}

The location of the target stars in the CMD
is consistent with the spectral types assigned in this work (Fig.~\ref{F:EBVHII}).
The three OBA stars are found in the upper part of the blue plume.
The red stars SD-5, SD-6 and SD-7
are enclosed in the vertical sequence of Galactic foreground
contamination.
SD-4, on the other hand, is above the tip of the RGB
consistently with the star being
a red-supergiant.

The loci of the youngest stars in the CMD is also consistent with
the theoretical isochrones \citep[Z=0.001 or $\sim$1/15\Zsun][]{MGB17}
but require a larger correction
than the reported \ebvfg=0.12 foreground extinction \citep[e.g.][]{Moal05}.
Internal reddening can indeed be expected, since
SD-1 to 4 are located in a concentration of neutral hydrogen \citep[][also B14]{YL97}.
The \ebv$\simeq$0.22 colour excess calculated from the Balmer decrement of SagDIG-HII\#3 
was adopted \citep{LGH03,TBS16}.

SD-3 is located at the main sequence
where isochrones of $ \log \left( age[yrs] \right) $=4.5--7.0
overlap, in accordance with its OB~V spectral type.
SD-2 is found 
between the $ \log \left( age[yrs] \right) $=7.0 and 7.49 isochrones,
as expected considering that BA--supergiants have evolved off the main
sequence and are slightly older than O--stars.
The location of SD-4 is consistent with a RSG with similar age $ \log \left( age[yrs] \right) $=7.49.

SD-1 is found between the
$ \log \left( age[yrs] \right) $=7.0--7.49 isochrones, 
however, it is more reddened
than the other programme stars.
If its \ebv=0.38 spectroscopic colour excess is taken at face value,
the isochrones would need an additional shift of
$\rm \Delta \left( F606W - F814W \right) \sim$0.2 mag and $\rm \Delta F814W \sim$0.3 mag
to account for the increased extinction.
This would bring SD-1 close to the $ \log \left( age[yrs] \right) $=6.49 isochrone (3 Myr),
that better agrees with the expected age of an O--type supergiant.

The unknown amount of internal reddening in SagDIG 
hinders the interpretation of the CMD
(SD-1,2,3 would be mistaken as intermediate$-$mass stars if \ebvfg~ was used)
and may hide a significant population of young and massive stars.
Extinction can explain the paucity of stars in the upper blue plume, which is
at odds with the extended emission of the galaxy in the UV.
These results may be considered a reminder that the total galactic stellar content
may be underestimated from the luminosity function,
and advices against selecting massive star candidates from
their location in classical CMDs only.

\section{The most metal-poor RSG to date}
\label{s:RSGZ}

Red supergiants are He--burning descendants of \Mini$\lesssim$30\Msun~ stars 
and the coolest stage in the life-cycle of massive stars.
They are gathering increased interest 
because some RSGs (9$<$\Mini$<$16.5--23\Msun) are the alledged
progenitors of type-IIP supernovae  \citep{Gral13}.
Besides, they are proving
reliable metallicity probes optimally reached by current,
powerful IR spectrographs: Keck--NIRSPEC, GTC--EMIR or VLT--KMOS
\citep[e.g.][]{DOK09,GKE15}.

Red supergiants are classically associated with M-types,
but 
the average spectral type shifts towards earlier types in metal-poorer galaxies
\citep{EFH85,LM12}.
This trend is beyond a classification bias
produced by the weakening of TiO bands when metallicity decreases,
and it is believed to reflect the fact that the Hayashi limit
shifts towards higher effective temperatures with decreasing metallicity
\citep[see discussion by][]{EFH85}.
For this reason, and because sometimes they exhibit spectroscopic variability, 
types as early as G0 are considered in the class of RSGs \citep[see discussion in][]{GF15}.

According to the latest compilation of Local Group RSGs
by \citet{LM12}
the most frequent spectral spectral type shifts
from M2~I in the Milky Way and the LMC, to mid-K types in the SMC and to
K0-1 in WLM (1/7--1/10\Zsun).
The RSG-candidate SD-4, with spectral type G0--2~I,
extends the spanned range of metallicity down to Z=1/21\Zsun~
and supports the metallicity dependence of the RSG spectral
types.

\section{Summary and conclusions}
\label{s:fin}

This paper presents the first spectroscopic census and confirmation of massive stars in SagDIG. 
GTC--OSIRIS optical spectroscopy has unveiled three OBA--type stars in the galaxy,
and one red supergiant candidate.
Pending the quantitative analysis of high-resolution optical and UV spectroscopy \citep[e.g.][]{Gal14}, these could be the most 
iron-poor massive stars of the Local Group.

The G0--2~I spectral type of the RSG-candidate SD-4 is
consistent with the reported trend of
earlier RSGs spectral types in environments of lower metallicity \citep{LM12}.
It extends the sequence down to Z=1/10--1/20\Zsun.

The four massive stars have been found in the Eastern part of the galaxy, 
matching extended UV emission, and close to the highest concentrations
of neutral hydrogen.
Their CMD location relative to theoretical isochrones is consistent with their spectral types
as long as  colour excess \ebv$\gtrsim$0.22, larger 
than the foreground value, 
is considered.
This demonstrates that internal reddening in SagDIG is not negligible, and
that the selection of candidate massive stars exclusively from their location at the optical CMD 
could dismiss a significant fraction of good candidates.

This paper opens the way to use SagDIG as a benchmark for studies of metal-poor massive stars.
The most urgent follow-up is high-SNR,  mid-resolution spectra
of the reported stars in order to confirm their low metal content.
The census of massive stars must also be completed,
and this requires deep multi-object spectroscopic observations.
In this respect, SagDIG is an ideal target for VLT--MUSE.
By providing spectra of all the stars
in the field of view,
MUSE will bypass the shortcomings of reddening 
and target-selection biases.
Observations are already planned.

\section*{Acknowledgments}

M. Garcia would like to thank F. Najarro for fruitful discussions and support from 
grants FIS2012-39162-C06-01, ESP2013-47809-C3-1-R and ESP2015-65597-C4-1-R.
Ricardo Dorda is thanked for his library of standard stars.
This research used NASA's Astrophysics Data System,
the SIMBAD database \citep{SIMBAD},
and the Aladin Sky Atlas \citep{aladin1,aladin2}.






\input{biblio}






\bsp	
\label{lastpage}
\end{document}

%% file: star_phot.tex
\begin{table*}
  \caption{ 
    Observed stars:
    spectral types (SpT) from this work, radial velocities (\vrad) used in Figs.~\ref{F:sp}--\ref{F:sg},
    and cross-identification numbers and photometry
    from  M14 and B14.
    \MV~ was computed from aparent magnitudes, distance modulus (SagDIG stars: DM=25.10 \citet{Moal05}; foreground stars: DM=14.50),
    and \citet{allen}'s intrinsic colors to estimate \evi~ and correct from extinction.
    The last column flags SagDIG membership.
  }           
\label{T:phot}      

\centering
\begin{tabular}{l l r | c l l l | l l l | r r l}
  \hline
ID    &  SpT      & \vrad  & M14       & F475W  & F606W  & F814W  & B14    & \textit{V} & \textit{I}       & \evi  & \MV & Memb.  \\
      &           & [\kms] & ID        &        &        &        & ID     &        &              &       &     &         \\
\hline                                                                                                       
 SD-1 &  Late--O~I&  -65   & 33466     & 20.189 & 20.125 & 20.024 & 8903   & 20.190 & 20.154       & 0.51 & -6.1 & yes   \\  
 SD-2 &  BA~I     &  -55   & 31722     & 20.355 & 20.278 & 20.120 & $\ddagger$  & 20.38  & 20.24   & 0.12 & -5.1 & yes   \\  
 SD-3 &  OB~V     &  -95   & 32939     & 21.408 & 21.480 & 21.515 & 8822   & 21.442 & 21.585       & 0.28 & -4.3 & yes   \\  
 SD-4 &  G0--2~I  &  -100  & 30481     & 21.558 & 20.503 & 19.394 & 8805   & 20.731 & 19.473       & 0.42 & -5.3 & yes   \\  
 SD-5 &  G4~III   &  -90   & 3019      & 20.887 & 20.055 & 19.144 & $\ddagger$  & 20.38  & 19.23   & 0.26 &  5.3 & fg    \\  
 SD-6 &  G5~V     &  -70   & $\dagger$ &        & 18.59  & 17.82  & 7783   & 18.875 & 17.912       & 0.07 &  4.2 & fg    \\  
 SD-7 &  K4~V     &  -50   & $\dagger$ &        & 18.76  & 17.74  & 8017   & 19.124 & 17.819       & 0.42 &  3.7 & fg    \\  
\hline
\end{tabular}

\noindent ($\dagger$)~ Star without M14-counterpart. F606W and F814W magnitudes estimated from B14's \textit{V}- and \textit{I}-magnitudes,
   using \citet{RGH05}'s transformations.\\
\noindent ($\ddagger$) Star without B14-counterpart. \textit{VI}-magnitudes estimated from M14, using \citet{RGH05}'s transformations.
\end{table*}